# Labor Contract Law

## -An Economic View

Yaofeng FU[*], Ruokun HUANG[**] and Yiran SHENG[***]

*Department of Finance, School of Economics and Management,*

*Tsinghua University, Beijing, P. R. China*

**Abstract:** China's new labor law – *Labor Contract Law* has been put into practice for over one year. Since its inception, debates have been whirling around the nation, if not the world. In this article, we take an economic perspective to analyze the possible impact of the core item–open-ended employment contract, and we find that it deals poorly with adverse selection, with moral hazard problems arise, which fails to meet the expectations of law-makers and other parties.

Version: Apr 2009

[*] Yaofeng FU, fuyf.06@sem.tsinghua.edu.cn, student ID 2006012405.
[**] Ruokun HUANG, huangrk.06@sem.tsinghua.edu.cn, rh382@cornell.edu, student ID 2006012402.
[***] Yiran SHENG, shengyr.06@sem.tsinghua.edu.cn, student ID 2006012400.

# Contents

# Introduction: Overall Debates

The *Labor Contract Law* (hereafter referred as the *Law*) in China is enacted in 2007 and became effective on the first day of 2008. And the idea of this new law was proposed in the year 2005. This new law enjoyed huge focus and popularity nationwide with media coverage for every single article within it. It was once reported that during the 28th Session of the Standing Committee of the 10th National People's Congress, which took place on the day of 30$^{th}$ June 2007, the Law was passed with 145 pro-votes and 0 against-vote, demonstrating a sharp contrast with laws and regulations passed before. Another example of its popularity is the fact that around 200,000 of pieces of suggestion were handed in within one month of open consulting from the public.

However, the popularity and majority concern cannot guarantee a perfect law. In fact, the *Law* has long been a subject of debate over the years passed since its inception.

The preliminary debate of this law comes from the fundamental concept within the law. The basic idea is to protect the "disadvantaged labor" against the "over-dominant" capital, which have made an indispensable contribution to China's economy since its open-up policy in 1979. On one hand, law makers argue that migrant workers in China have never got the chance to protect their own rights since the open-up, i.e. they got little symbolic "compensation" for from the employers if they were hurt or even disabled during their work, and even got nothing back at all, which is often the case, because they are not officially hired by their employers – no contracts, no wage guarantee, not to mention the insurance. On the other hand, economists argue that the law leans towards the employees, which will results in the consequence of increasing the labor costs of companies, thus making China losing the low-labor-cost advantages, which has attracted tremendous amounts of investment from abroad. As for economists, they believe that a better social welfare system is a more practical way to tackle with the problems facing migrant workers, rather than a change of law. And if necessary, the change of law should not deviate from the stand of protecting both sides simultaneously.

The authors of this article agree with the latter point of view, not simply because we are the potential economists in the future, but further, evidence shows that the leaning-towards workers policy is not favorable to all the workers as a whole, but only a small proportion of employees, who have already been quite well-off because of their already highly accumulated human capital. For those in lack of human capital or poor in skills, as in the case of most migrant workers in China, this policy can create unexpected consequences for those workers as law-makers cannot anticipate. The underlying mechanism lies in the fact that the over-protective items in the *Law* make the employers more cautious and hesitant when deciding on whether or not to hire a person, especially in the cases where the job candidates are at the margin of being accepted due to their not-that-high skills and experiences, making those, for example, migrant workers lose the potential opportunities of employment otherwise they would not lose. To summarize, the ideal of those law-makers may appear noble, i.e. to protect "disadvantaged labor", however, the impact may well be the opposite to their intention due to their lack of knowledge of economics of organizations.

Nowadays, the debate concentrates largely on two other issues instead: should the law in China coerce employers to sign an open-ended employment contract with their employees or not, and additionally, should those employers pay a minimum wage for the workers. The specific items in the law are displayed as follows:

Article 14 (open-ended contract): "If an Employee proposes or agrees to renew his employment contract or to conclude an employment contract in any of the following circumstances, an open-ended employment contract shall be concluded, unless the Employee requests the conclusion of a fixed-term employment contract."
Article 20 (minimum wage rate): "The wages of an Employee on probation may not be less than the lowest wage level for the same job with the Employer or less than 80 percent of the wage agreed upon in the employment contract, and may not be less than the minimum wage rate in the place where the Employer is located."

The debate between law-makers and economists centered much on the open-ended employment contract, which is the core of the new *Labor Contract Law* in China, and also the main achievement that law-makers are proud of. Both sides are using seemingly well-grounded arguments to support their points. The law-makers argue that the status quo of the labor-capital relation should be re-established because the workers in China share a "natural" disadvantage over their employers because of two reasons, one is the lack of effective labor union (in fact, the union in China mainly concentrates on the internal affairs in the organizations, rather than protecting the legal rights of the workers); the other reason lies in the poor welfare system in China, leading to the lack of proper compensation and protection against possible lay-off. The open-ended employment contract, they argue, can solve these problems by offering better protection to workers against this situation, and they often quote the increase of the contract-signing percentage from less than 30% to 93% in the last year as an evidence of benefits from the open-ended contracts. On the other side, the economists, however, believe that this design of law twisted the incentive of employers and employees as well, making the former less willing to hire and the latter more "eager" to protect their legal rights, resulting in the consequence of increasing unemployment. This argument cannot be verified yet because the current financial crisis can contribute a lot towards the rise in unemployment rate.

The debate of minimum wage rate among the two parties shows the similar logic of reasoning, which involves the law-makers intended to protect the workers, in probation or after being fired, while the economists criticized it because of the twisted incentive problem.

In conclusion, authors of this article believe that the *Labor Contract Law*, especially the articles in which open-ended employment contracts and minimum wage rate, has brought more harm than benefits to the workers involved, and the reasons for the argument are as follows: the costs of employment for companies increase, making them unwilling to hire; on the other side, the workers have no high incentive to accumulate their human capital, rather, they will focus on taking legal actions against employers. Both of these will result in the inflexibility of employment and unemployment, especially for the "disadvantaged" groups, such as migrant workers, women, new graduates from college etc., who are supposed to be protected, rather than harmed.

# Economic Evaluation in Organizational Theory Approach

One of the elementary goals of the new labor contract aims at is making businesses responsible for fairness and equality in the economy. This is questioned, if not totally rejected, by modern economic theories. Businesses' key issue is productivity, inefficient businesses will reduce benefits or close their doors entirely, an outcome undesirable to the whole society as it leads to higher unemployment and very likely, greater unfairness.

Various theoretical approaches can be taken to address such issues in detail, this article focuses on those based on the organizational theories we have studied in this course. In specific, we address the informational problems arise from the new restrictions regarding the labor contract terms the *LAW* imposed on businesses. Adverse selection and moral hazard issues are analyzed in the following paragraphs.

Our topic derives from Chapter 3, Article 14, in the *LAW,* which is stated as follows:

> An Employer and an Employee may conclude an open-ended employment contract upon reaching a negotiated consensus. If an Employee proposes or agrees to renew his employment contract or to conclude an employment contract in any of the following circumstances, an open-ended employment contract shall be concluded, unless the Employee requests the conclusion of a fixed-term employment contract:
>
> (3)Prior to the renewal, a fixed-term employment contract was concluded on two consecutive occasions and the Employee is not characterized by any of the circumstances set forth in Article 39 and items (1) and (2) of Article 40 hereof.

Productivity is the critical concept to understand businesses' behavior, thus making information structure of workers' productivity very essential while discussing the effectiveness of this item in the *LAW.* In the following analysis, we use traditional informational economics concepts, i.e. adverse selection and moral hazard to emphasize possible market distortions the *LAW* would potentially bring. Then, we will discuss a few additional topics regarding macroeconomic conditions and international experiences, which are quite important in evaluating the *LAW*, yet have relative little relevance to our course.

# Section I A simple model: The methodology of firing employees

We are going to give an explanation of how firms fire workers. We claim that firm cannot observe the true productivity of workers in single period, as what was assumed in the model in class. We assume that the observation process need more than one period, by the following model:

**Assumption 1**

Firm cannot tell the exact productivity of a worker, but can confirm that (100% confidence level) the productivity are in a particular interval, or

$$P(\theta_i \in [a,b]) = 1$$

**Assumption 2**

The firm at each period can only discover the relatively some worst workers (in some interval), or

We can surely distinguish from $\begin{matrix} \theta_i \in [a,c] \\ \theta_i \in [c,b] \end{matrix}$.

**Assumption 3**

The firm fire workers, or the workers quit, when the firm discover the worker $\theta_i \in [a,c]$ and that $c < \bar{\theta}$.

**Assumption 4**

The workers are uniformly distributed from $[\theta_l, \theta_h]$. The firm at period t can only observe surely $\theta_i \in \left[\theta_l + \frac{t}{n}(\theta_h - \theta_l), \theta_l + \frac{t+1}{n}(\theta_h - \theta_l)\right]$ or surely $\theta_i \in \left[\theta_l + \frac{t+1}{n}(\theta_h - \theta_l), \theta_h\right]$. Thus it needs n periods to entirely discover the workers' productivity.

**Object:** Firm wants to hire workers with high productivity.

**Model:**

When there are n periods, there is one worker work for the firm. If the firm discovers the employee's productivity falls in the dark area, he would be fired, in corresponding period.

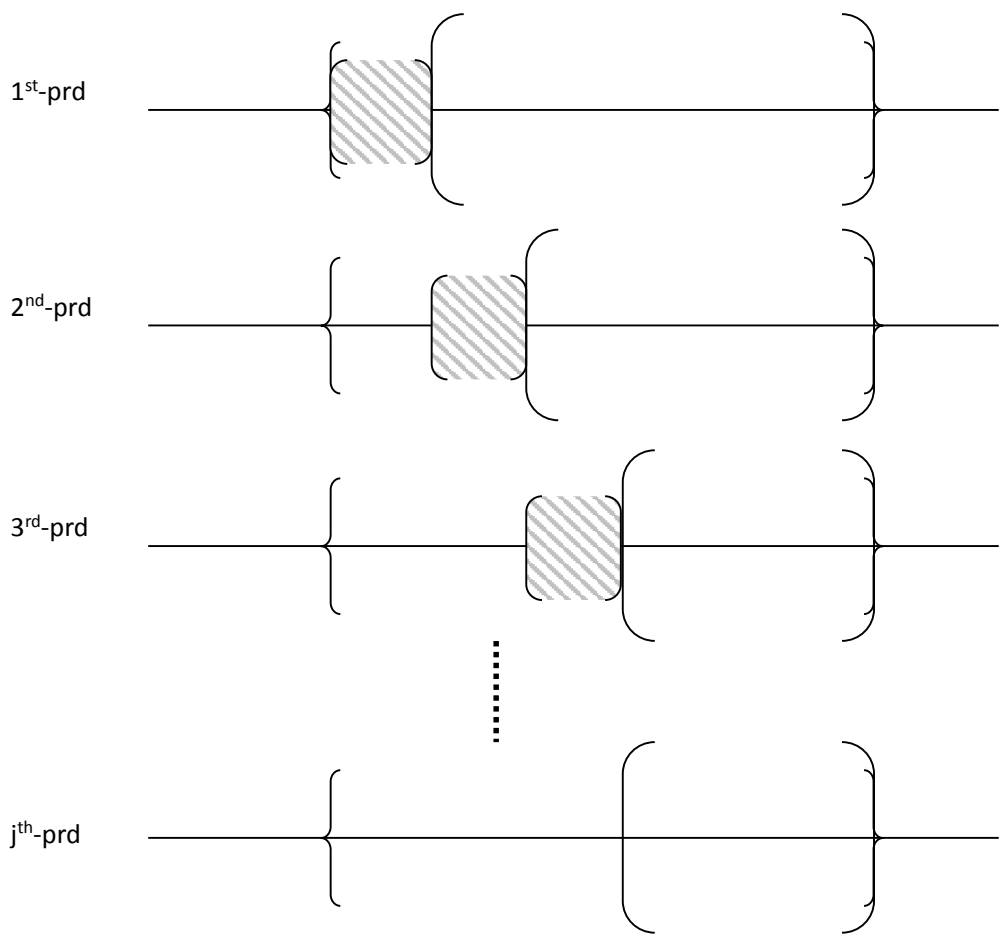

The firm fire unqualified workers, whose productivity is less then market average, step by step from 1st period to jth period, where the remaining workers' productivity are all above market average.

When there are less than n periods available, say m periods, for a firm to justify and fire workers,

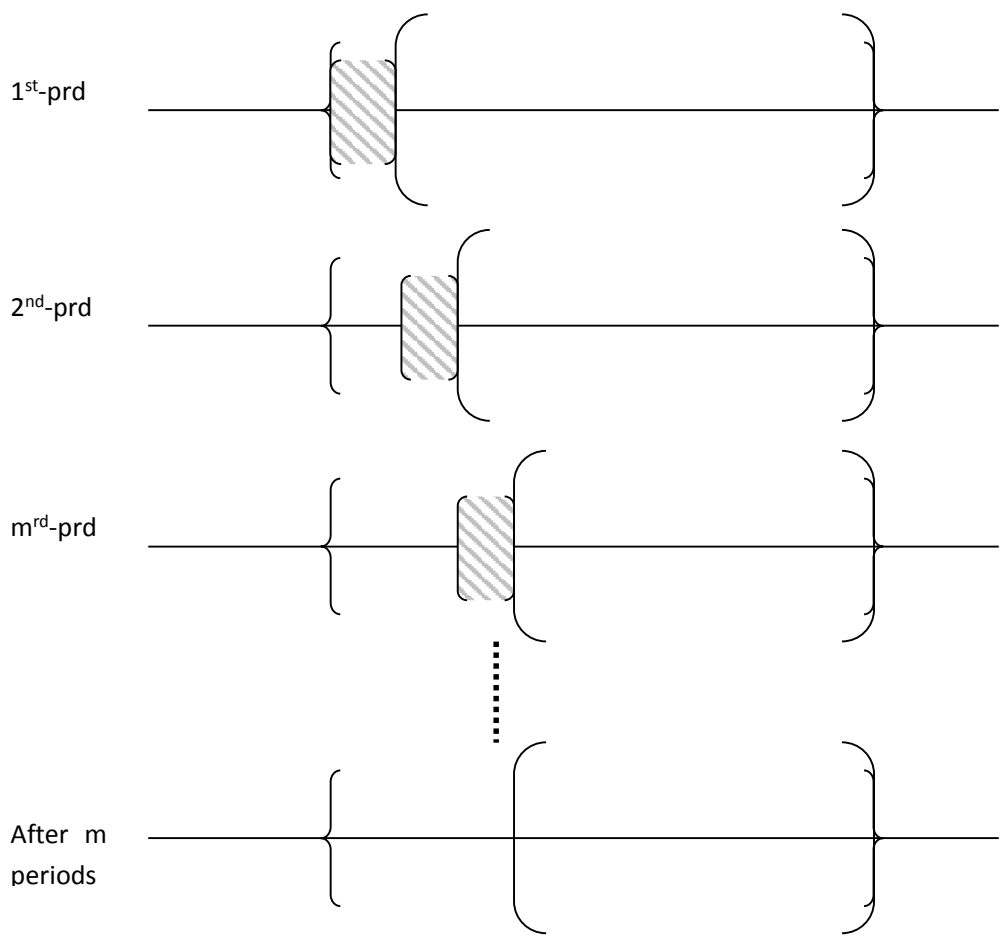

1st-prd

2nd-prd

mrd-prd

After m periods

Firm cannot distinguish which worker's productivity is above average, so there is some probability that a worker in the firm is worse than average, under the condition that

$$\frac{m}{n} < \frac{1}{2}$$

Adaptability of Model:

The assumption of the model is similar to what firms in reality do when they consider to fire workers. The key point is that it is easier to find a work with low productivity rather than high productivity. Bosses are more likely to be convinced that one employee is unqualified. When a worker does one thing wrong, it will be a long-lasting negative impression.

In the law, m is set to be 3. Suppose our model is reasonable, we can draw a conclusion that if the total periods required for fully assessment is less or equal to 6, the law will make negative contribution to information transparency.

However, in reality the firm can never be '100%' sure about a judgment. Since this increase the uncertainty in firm's decisions, the critical value m* in our model seems to be underestimated. So, the critical number of period for fully assessment of 6 may be overestimated. Five or four might be suitable as a critical value for n*, which is fairly reasonable in reality.

**Implications:**

1. Firm gathers information of workers step by step. The more periods it has, the more accurate it can tell about the worker.
2. Limited periods sometimes unable firm to gather information of the worker. If there are less than half of n periods (required to reveal complete information), the firm has possibility that its worker's productivity is lower than average.

## Section II adverse selection

Adverse selection is common in a labor market. The employer does not have the information about a worker's productivity in general before hiring and test him for a while. Also, there is information asymmetry between a worker's current employer and his potential employers if he quits or gets fired. Thus, we view the adverse selection as a three-pronged interaction among an employee, his current and potential employers. The firm will presumably try to prevent turnovers among its better workers, while firing the worse components of its workforce. (We do not distinguish firing and quitting for the firm can offer the worker a period wage lower than his market value, thus making him quit voluntarily) It could significantly impair a worker's freedom to change jobs and reduce his pay compared to what he can get under perfect information. (Greenwald, 1986) Such effects are weakened if the firm is allowed to fire workers freely, by creating labor market division, meaning group workers of different skills level into different submarkets (these markets are denoted second-hand markets, the markets of workers who get fired after certain periods). That is due to the option value of firing.

One very important limit the *LAW* imposed upon firms is, for how many periods can a firm fire worker freely. The number is set to 2 by law, a employer can fire worker within the first two periods, however, if the contract enters a third period the firm has to sign a lifelong contract. We argue that the number of periods before a long-term employment relationship is critical in dealing with the adverse selection problem.

The adverse selection problem in labor market and the equilibrium in second-hand market is discussed by (Akerlof, 1970) and (Greenwald, 1986). Our analysis basically follows their framework.

**The model**

Under symmetry information, each worker is hired and paid according to his real productivity, that is $w_i = \theta_i$, the former term denotes his wage, and the latter term his productivity.

Under asymmetry information, we build up three models with the different periods within which firms could fire workers freely.

# One period model

This can be viewed as a case no firing option is offered. The firm faces a pool of labor and hires randomly from the pool. Once hired, the worker need not worry about being fired. Here are the assumptions.

a. Each worker in the pool is characterized by an unobservable productivity $\theta_i$, the worst and best worker in the market have $\theta_L$ and $\theta_H$

b. Denote $N(\theta)$ a continuous function of numbers of people who has productivity $\theta$

The average quality of the market is given by $\bar{\theta} = \dfrac{\int_{\theta_L}^{\theta_H} \theta N(\theta) d\theta}{\int_{\theta_L}^{\theta_H} N(\theta) d\theta}$

In a competitive market the firm offers any worker it hires a wage level:

$$w^* = \bar{\theta}$$

Indicating a zero-profit firm. The firm will not hire anyone if θ-bar is negative, meaning the whole market collapses.

# Two-period model[1]

The firms hire people from the entry level market with all workers look identical. After one period of employment relationship, each employer finds out the real productivity of its employees. At the end of period one, workers are offered 2 options. They can either stay with their current employers or change jobs. In the latter event, they enter a second hand market.

---

[1] This model is adjusted from Greenwald's paper Adverse Selection in Labor Market 1986.

We carry out the same assumptions from 1-period model, to describe the basic structure of entry level market. Addition assumptions regarding the second-hand market are as follows:

**Assumption c:** Employers in the second-hand market have no more information other than the workers' employment history to infer the productivity of its potential workers.

**Assumption d:** The firm decides which employee to keep based on his productivity. That is given the equilibrium wage $w^*_1$ in the second-hand market, a worker's quitting decision is characterized by the following function:

$$q_i = \text{probability of worker i changing job} = \begin{cases} \mu(\theta_i > w^*_1) \\ 1(\theta_i \leq w^*_1) \end{cases}$$

$0<\mu<1$ is a random quit factor, which is assumed to be exogenous. We will discuss this assumption later both in 2-period model and the 3-period.

**Assumption e:** Firms are restricted to single period, non-contingent wage offer. For work i, the firm's second period wage offer is:

$$w_i = \begin{cases} w^*_1 (\theta_i > w^*_1) \\ 0(\theta_i \leq w^*_1) \end{cases}$$

The worker receives his market value inside the firm. Detailed proofs can be found in appendix 1, (Greenwald, 1986), we take it as an assumption in later analysis.

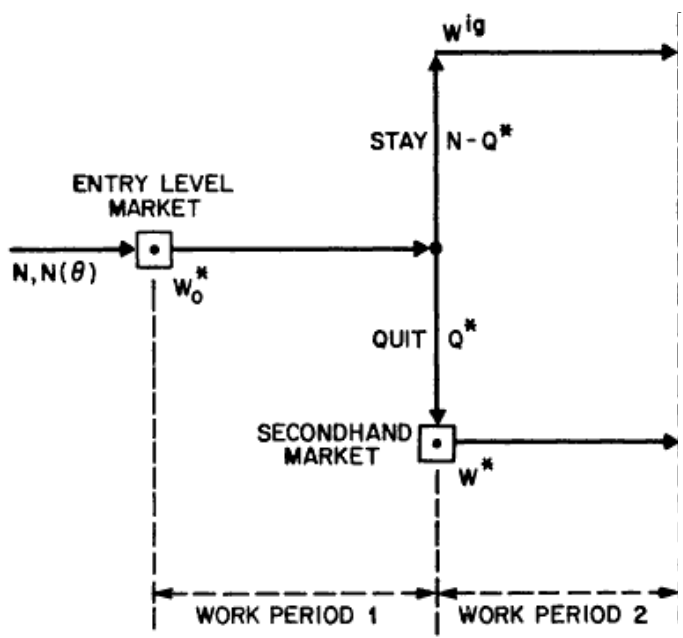

FIGURE 1
Two period model

(This graph is adapted from Greenwald's version with minor modifications.)

**Solving for equilibrium wage in second-hand market.**

The equilibrium condition in second-hand market implies that the market wage level equals to the average productivity. (The result from one period model)

$$\overline{\theta_1} = w^*_1$$

And $\overline{\theta_1}$ in turn depends on $w^*_1$, the critical value of productivity the first period employer used to make firing decisions.

$$\overline{\theta_1} = M(w^*_1)$$

Where M(w) is defined as, the market contains all workers from entry level market with productivity [$\theta_L$, w] and proportion μ of workers with productivity [w, $\theta_H$] from the entry level market.

$$M(w) = \frac{\int_{\theta_L}^{w} \theta N(\theta)d\theta + \int_{w}^{\theta_H} \mu\theta N(\theta)d\theta}{\int_{\theta_L}^{w} N(\theta)d\theta + \int_{w}^{\theta_H} \mu N(\theta)d\theta}$$

Solving the equation $w^*_1 = M(w^*_1)$, we have the equilibrium wage level in second-hand market, as is shown in the graph. Notice in the graph θ_L is assumed to be greater than zero, there for $M(0) = \frac{\int_{w}^{\theta_H} \mu\theta N(\theta)d\theta}{\int_{w}^{\theta_H} \mu N(\theta)d\theta} = \bar{\theta}(>0)$, the average quality in entry level job market. And $M(\infty) = \frac{\int_{\theta_L}^{\theta_H} \theta N(\theta)d\theta}{\int_{\theta_L}^{\theta_H} N(\theta)d\theta} = \bar{\theta}$. A more general case allowing θ_L < 0 would look a little bit different graphically.

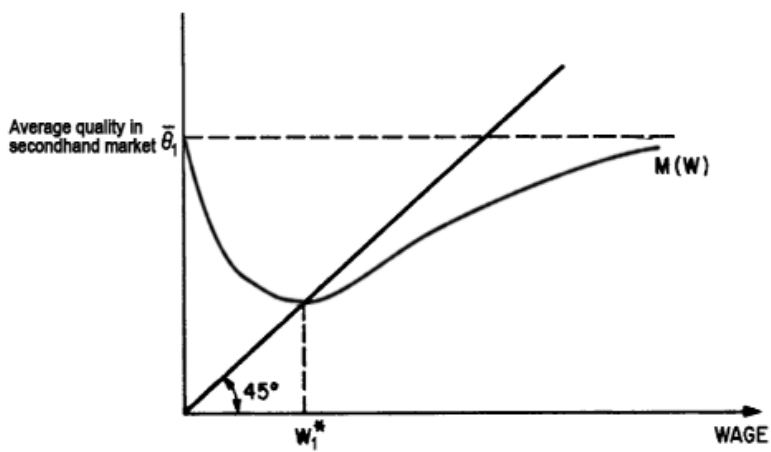

FIGURE 2
Equilibrium secondhand market

Notice $w^*_1 < \bar{\theta}$

What's more, the assumption μ>0 here is necessary, since when μ=0, we easily find out M(w)<w for any w>0. The only possible equilibrium wage level is $w^*_1 = 0$. This means the second-hand market is one that looks like a market we mentioned before in one-period model. Or in other words, the second-hand market collapses. The fired workers are those with negative productivity and no employer would be willing to hire then at any positive wage level, these works have but one choice- go home and do nothing. Random quitting is essential for second-hand market to exist.

**Solving for entry level market equilibrium wage**

The equilibrium condition for this market is a zero profit condition for two periods since firms are competitive to the same pool of labor. That is:

$$\sum \theta_i - Nw^*_0 + \sum \theta_j - Qw^*_1 = 0$$

Where $N = \int_{\theta_L}^{\theta_H} N(\theta)d\theta$ denote the number of workers in the entry level market, and $Q = (1-\mu)\int_{w^*_1}^{\theta_H} N(\theta)d\theta$ denotes the number of workers remaining in the firm (or more precisely the firms), and $\theta_j > w^*_1$ for all j. Take expectation of both sides, we have

$$N(\bar{\theta} - w^*_0) + Q((\int_{w^*_1}^{\theta_H} \theta N(\theta)d\theta)/(\int_{w^*_1}^{\theta_H} N(\theta)d\theta) - w^*_1) = 0 \quad (*)$$

$\bar{\theta}_2 = (\int_{w^*_1}^{\theta_H} \theta N(\theta)d\theta)/(\int_{w^*_1}^{\theta_H} N(\theta)d\theta)$ is the average quality of workers left in the firm. Solving the equation, we get $w^*_0$.

Since the second term in (*) is greater than zero (the workers remaining in firms have higher average productivity than those who quit, i.e. the average quality of second-hand market), we conclude,

$$w^*_0 > \bar{\theta}$$

**Conclusion**

Compare $w^*_0$, $\bar{\theta}$, $w^*_1$, we have:

$$w^*_1 < \bar{\theta} < w^*_0 < \bar{\theta}_2$$

The four terms stands for the equilibrium wage in second-hand market, the average productivity in entry level market, the equilibrium wage in entry level market, the average productivity of those who stay with their current employers.

The special features of two period firing options compared to the no firing situation which is the case of our one-period include:

1. For an entry level market with $\bar{\theta}<0$, the one period model suggests there will be no market at all, the workers with positive productivity as well as the firms will suffer. Within the framework of 2-period model, on the other hand, the firm will hire from such an negative expected productivity market as long as the firm can fire those bad workers at the end of period 1 and keep the good workers and pay them the same as (perhaps a little bit more) the bad workers' wage in the second-hand market, in this case it is 0 ($w^*_1<\bar{\theta}<0$). There will exist a positive equilibrium wage $w^*_0$ in the entry level market, sufficiently large over $\bar{\theta}$. A simplified version of this model is provided in Chapter 3 of our textbook of this course.
2. Since $w^*_0$ is driven up by market competition, there will be free riders in the entry level market. The bad workers enjoy a high 1st period wage at the expanse of shrank pay of the good workers. The good workers are compensated in period 1, but their pay is significantly undermined in the 2nd period, for they are not able to prove their true productivity in the second-hand market, thus trapped with their current employer and receive a low pay ($w^*_1$) compared to their high productivity($\bar{\theta}_2$).

The two-period and 1st period end option of firing manage to solve some problems but still, the fundamental issues of adverse selection remain unresolved. What would be a better solution? A 3-period needs to be forged to answer this question. Now we know the 2-period case provide the first-degree division of labor markets by creating a second-hand market, and distinguish the bad workers from the entire work force. If employer is allowed to fire workers at the end of period 2 before entering into a period 3 (a long-term contract), the labor market will be divided further into 3 second-hand markets, and information regarding workers' real productivity is better revealed and classified. This regime is what the *LAW* offers to the economy while the old system is even more flexible with infinite periods and free firing option at the end of each period.

## Three-period model

The model follows the same set of assumptions in the previous discussion. In addition, one more assumption is required here:

**Assumption f**: Firms have perfect information of a worker's employment history. [2]

Mathematical notations are carried out as follows.

$\bar{\theta}$        the average productivity in entry level market

N(θ)      the number of workers with productivity θ in entry level market

N         The number of workers in the entry level market

$N_1, Q_1$    The number of workers in the second market, and the number of workers who remain with their current employers the end of the 1st period

$N_2, Q_2$    The number of workers in the third market, and the number of workers who stay with their current employers the end of the 2nd period

$N_2', Q_2'$   The number of workers in the double second market, and the number of workers who stay with their current employers the end of the 2nd period

$w^*_0$      The equilibrium wage level in the entry level market

$w^*_1$      The equilibrium wage level in the second market

$w^*_+$      The wage the workers receive who choose to stay with their current employer after period 1, notice $w^*_+$ and $w^*_1$ don't have to be equal

$w^*_2$      The equilibrium wage level in the third market

$w^*_2{}'$     The equilibrium wage level in the double second market

Note the second market means the labor market for laid-off workers at the end of the first period, and the third market is the market for laid-off workers at the end of the second period. And the double second market is the market of twice fired workers at the end of the second period.

---

[2] Here is our major distinction compared to Greenald's work (1986), which assumes all laid-off workers just enters into a same second-hand market. However, assumption f in our model implies different workers with different employment history are automatically grouped into four sub-markets based on the whether and when they change their job. The four markets and its definition can be found on the next page.

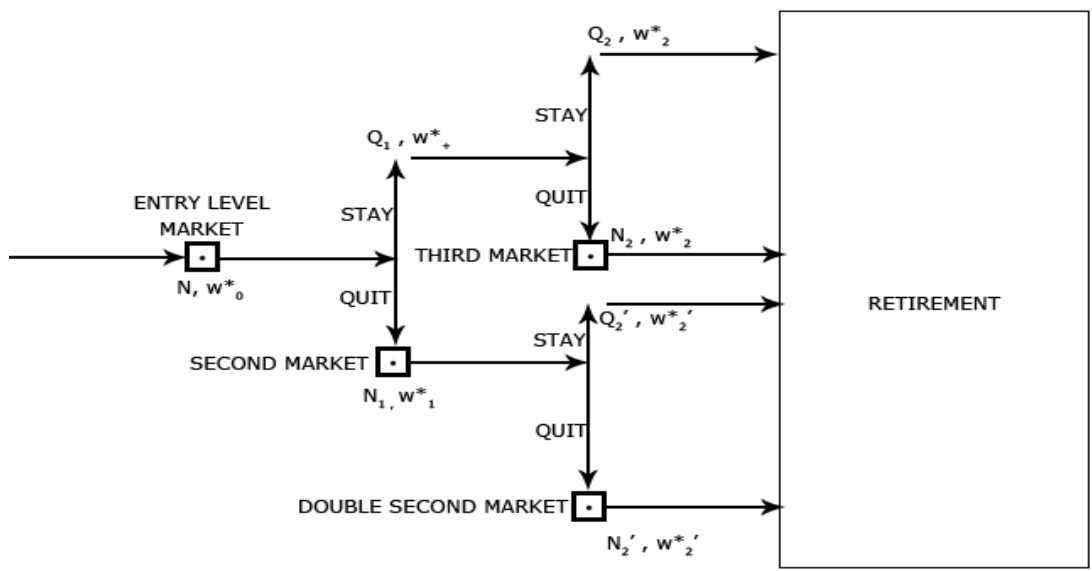

(Note this figure is significantly different from Greenwald(1986)'s version.)

Some basic relationships between these variables are:

$$N = \int_{\theta_L}^{\theta_H} N(\theta)d\theta$$

$$N_1 = \int_{\theta_L}^{w^*_+} N(\theta)d\theta + \int_{w^*_1}^{\theta_H} \mu N(\theta)d\theta w^*_+$$

$$Q_1 = \int_{w^*_+}^{\theta_H} (1-\mu)N(\theta)d\theta$$

$$N_2 = \int_{w^*_+}^{\theta_H} (1-\mu)N(\theta)d\theta - \int_{w^*_2{}'}^{\theta_H} (1-\mu)N(\theta)d\theta$$

$$Q_2 = \int_{w^*_2}^{\theta_H} (1-\mu)N(\theta)d\theta$$

$$Q_2{}' = \int_{w^*_2{}'}^{\theta_H} (1-\mu)N(\theta)d\theta$$

$$N_2{}' = \int_{\theta_L}^{w^*_+} N(\theta)d\theta - \int_{w^*_2{}'}^{\theta_H} (1-\mu)N(\theta)d\theta$$

Following the same logic while deriving the equilibrium wage in the secondhand market of our 2-period model, we conclude the equilibrium wage in 3rd market and double second market equals to the average productivity in these market because the markets yield a last period employment contract and no future value of quitting or firing value needed to be reflected in the market wage.

The mathematical interpretation is:

$$w^*_2 = M_2(w^*_2)$$

$$M_2(w) = \frac{\int_{w^*_+}^{w^*_2} \theta N(\theta)d\theta + \int_{w}^{\theta_H} \mu\theta N(\theta)d\theta}{\int_{w^*_+}^{w^*_2} N(\theta)d\theta + \int_{w}^{\theta_H} \mu N(\theta)d\theta} \quad : (3\text{-}1)$$

$$w^*_2{}' = M_2{}'(w^*_2)$$

$$M_2(w) = \frac{\int_{\theta_L}^{w^*_2{}'} \theta N(\theta)d\theta + \int_{w}^{\theta_H} \mu\theta N(\theta)d\theta}{\int_{\theta_L}^{w^*_2{}'} N(\theta)d\theta + \int_{w}^{\theta_H} \mu N(\theta)d\theta} \quad : \quad (3\text{-}2)$$

However, sadly enough, with these two equations we can't solve the two equilibrium wages, since the equations depend on $w^*_+$.

Now go back to the end of first period, for a worker is qualified to stay inside the firm (i.e. a random quitter), if he choose to quit his life time wealth in the following periods is $W_q = w^*_1 + w^*_2{}'$, if he choose to stay in the firm his lifetime wealth in the following periods is $W_s = w^*_+ + w^*_2{}'$. The firm's offer $w^*_+$ to him must essentially make him indifferent between $W_s$ and $W_q$. Thus, we have:

$$w^*_1 + w^*_2{}' = w^*_+ + w^*_2{}' : \quad (3\text{-}3)$$

For simplicity, we do not introduce any discount factor here, the impact of a discount factor is unimportant and will not alter the final solution dramatically.

For entry level market equilibrium wage, the condition is a zero profit condition as discussed earlier. That is:

$$N(\bar{\theta} - w^*_0) + \int_{w^*_+}^{\theta_H}(1-\mu)\theta N(\theta)d\theta - Q_1 w^*_+ + \int_{w^*_2}^{\theta_H}(1-\mu)\theta N(\theta)d\theta = 0 \quad (3\text{-}4)$$

For the firm who starts hiring from the beginning or the second period, a zero profit constraints also applies:

$$\int_{\theta_L}^{w^*_+} \theta N(\theta)d\theta - Q_1 w^*_+ + \int_{w^*_+}^{\theta_H} \mu\theta N(\theta)d\theta - N_1 w^*_1 + \int_{w^*_2}^{\theta_H}(1-\mu)N(\theta)d\theta - Q_2 w^*_2 = 0 \quad (3\text{-}5)$$

Theoretically, solving (3-1) to (3-5) gives us $w^*_0$, $w^*_1$ $w^*_+$, $w^*_2$, $w^*_2$, however the results as well as the solving process are complex and unenlightening, we will not list them here. Instead some intuitions lie behind are stated here.

**Proposition (3-1)**   $w^*_2{'} < w^*_2 < \bar{\theta}_{Q2}$

At the end of the second period, the workers are divided into 2 groups, and 4 types of employment status. The equilibrium wage level in the third market is larger than that of the double second market. $\bar{\theta}_{Q2}$ denotes the average productivity of workers still left with their initial employers after 2 periods and enters into a permanent contract. This is true because the stream of workers who quit at the end of the first period has a lower average productivity than those choose to stay. And this piece of information can is reflected in the market, by labor market division, the third market and the double second market automatically distinguish them. However, the best workers in the entry level market (those who stay with one employer all the way), are still underpaid compared with their real productivity. The division of labor market can achieve a second-degree productivity division, yet still not revealing the full information. The employers can still prevent the labor turnover among their best employees. These good workers can tell the market they are not the worst kind (this is something they cannot achieve in the 2-period model), but they still cannot tell that they are the best.

**Proposition (3-2)**   $w^*_+ < w^*_2{'} < \bar{\theta}_{Q1} < w^*_1$

First, it's interesting to realize $w^*_+ < w^*_1$, those who choose to stay with their current employer receive a lower pay for the following period than those who leave. This is true, because those who stay have better compensates in the third period, therefore they are willing to suffer a temporary opportunity cost. The $w^*_2{'} < \bar{\theta}_{Q1} < w^*_1$ part follows the same line of reasoning as the 2-period model. What is critical about the payments in the second period is this: $w^*_+ < \bar{\theta}_{Q1}$ ($< \bar{\theta}_{Q2}$ furthermore), indicating the workers staying with a same employer not only suffers in the third period, but also are underpaid in the second period, while the lousy workers are overpaid ($\bar{\theta}_{Q1} < w^*_1$), in terms of their true productivity.

**Conclusion**

To summarize, the internal mechanism of labor productivity division has lagged effects upon wage payments. After distinguish good and bad workers, the employer has no incentive to offer the good one better pay consistent with his real productivity, since the worker cannot prove himself so good if he quit, and thus receives a wage outside measures the average productivity of a pool of workers worse than him. The worker has to wait one more period to enjoy a wage rise (only if the addition period exists). The 2-period mechanism achieves a 1st-degree productivity distinction and 0-degree wage distinction. The 3-period mechanism achieves a 2nd-degree productivity distinction and a 1st-degree wage distinction. If we were to build up a n-period model, the results will be the same. The adverse selection problem is dealt better if more periods are offered, but never perfectly. The good workers are better when period's numbers increase, but they are always underpaid, only the gap becomes narrower. Information advantage a current employer possesses gives them the ex. post power to undermine good employees' wage, thus benefits in a sense of lower production costs. This in not parato optimal, obviously. However, ex. ante competition among employers tends to drive up the upper branch of labor market equilibrium wage. The relative bad workers receive a better pay at the cost of a steep under-payment of the good workers in terms of productivity. The first best solution of perfect information case is never feasible, but can be approached through the increasing of period's numbers.

## Challenges and further works to be done

1. The only dimension we used to compare 2 and 3-period mechanisms is the degree of divisions of productivity and wage payments. We say any finite periods mechanisms cannot achieve a first-best (Pareto optimal) solution, with only limited improvements offered. However, the welfare improvements is not provided in the above theory, i.e. compare $w^*_0 + w^*_2 + w^*_+$ of the 3-period model with the $w^*_0 + w^*_1$ of the 2-period model. This is not insignificant in any case, but due to our limited mathematical skills in solving (3-1) to (3-5), such comparison is not provided. Further study on this issue should be carried out.
2. The assumption of random quit is not valid as period's number increases. We treat the quit rate in different periods as fixed and exogenous, while in reality the rates could differ significantly across periods. Also, the exogenous assumption is in question; a better approach should treat $\mu$ as an endogenous variable, a variable the work force as a whole could choose to maximize their utility. The change could have significant impact upon the size of different divisions of labor markets.
3. The concept of divisions of labor market itself is not fully consistent with reality. By some easy calculation, we found the number of sub-markets is $2^{n-1} - 1$ for n-period model, increasing at a geometric rate. The transaction costs associated with such an incredible rate is dramatic. Thus it is quite reasonable to doubt the actual effects we expected to see of labor markets division as a well the likelihood of such a delicate division happens.

Based on these challenges, it is almost certain that an infinite periods approach will not bring about too much exciting solution; for one thing, it fails to achieve a parato efficient outcome, for another, its theoretical effectiveness is quite limited due to unrealistic assumptions.

## Section III Moral Hazard Problem

We will use the model from (Holmstrom, 1979) to show that effort level is not efficient.

In this case Pareto-optimal sharing rules s(x) are generated by the program:

$$\max_{s(x),a} E\{G(x-s(x))\} \tag{1}$$

$$subject\_to\_E\{H[s(x),a]\} \geq \overline{H} \tag{2}$$

$$a \in \arg\max_{a' \in A} E\{H[s(x),a']\} \tag{3}$$

Where the notation "argmax" denotes the set of arguments that maximize the objective function that follows. G(·) stands for the utility function for firm, while H(·) stands for utility function for worker. $\overline{H}$ is the reservation utility of worker. x is output level, a is effort level and s(x) is wage. The *first-best solution* is the said to be the solution subject to only (2), ignoring (3), which differs from the *second-best solution*, subject to (2) and (3).

From the Corollary 1 in (Holmstrom, 1979), *first-best solution* and *second-best solution* are different if $f_a(x,a) \neq 0$. This is very likely to be true, or in other words, effort does influence the output level. Moreover, normally speaking $f_a(x,a) > 0$, thus *second-best solution* is greater than *first-best solution*, which means given wage, the effort will be lower than optimal level.

In conclusion, the effort level is not optimal. More specifically, given wage, effort level is lower than optimal.

# Policy Implications and International Experience

The open-ended employment contract introduced by the *LAW*, fixed the free trial periods number to two, within which the employer can make hiring and firing decisions independently. The key point here is to protect the workers' rights while allowing the firms some degree of freedom to hire and select good workers and fire those with relatively low productivity: a tradeoff between efficiency and equality.

However, as we argued, with adverse selection problems in presence, the number of periods the employers have allowing for free employment decisions is critical in achieving the goal of efficiency, namely reducing adverse selection to an acceptable level. The procedure introduced by the *LAW* is characterized by our 3-period model in section II. And a 2-period free trial is definitely not enough. Not only in a sense that adverse selection is merely roughly dealt (with one $1^{st}$-degree of compensating identification accomplished), but also in a manner equality is promoted poorly (significant underpay for the high-end workers). In fact, as we argued, the inequality arises in labor contract is preliminarily caused by the adverse selection, i.e. the employer is able to undermine the compensation to its best workers, who cannot prove their real productivity to the market due to information asymmetry. In this regard, eliminating adverse selection and promoting equality become a same goal, yet the solution offered by the *LAW* deals with both poorly.

Furthermore, we pointed out the old mechanism existing prior to the *LAW*, offering infinite intervals of free trial, also has its limitations. First of all, it cannot eliminate adverse selection and achieve the first-best solution in perfect information case, however large the period's number n is. Also, the effectiveness of increasing period's number declines dramatically due to factors we mentioned at the end of section II. Therefore, returning to the old *Labor Contract Law* is also not a good idea. If period's number prior to open-ended contract is the only policy tool available to us, the optimal number should be worked out carefully rather than set to a roughly chosen one.

Also we took into consideration of potential moral hazard problems. The workers' incentive to diligent working is weakened, if not significantly distorted, since the fear to lose jobs resolves once they enter into an open-ended contract and the employer cannot fire them at will. Unlike adverse selection, which generally exists before the introduction of the *LAW,* this problem is exactly newly brought up the *LAW.* As there is no ideal theoretical mechanism we can adapt to deal with moral hazard, we conclude this is one of the unnecessary costs the *LAW* generated with no evident benefits created at the same time.

Therefore, we are confident to say the *LAW* failed to meet its initial expectation and to satisfy its original goodwill.

In an international perspective, we found Germany and India once had laws issued with similar content written inside, imposing constraints on firms' hiring and firing decisions. Countries with these kinds of laws—Germany, Italy, France, India and some parts of Canada—bear higher unemployment rates, around nine and ten percent. This does not include workers leaving these countries and those not registered, so the true number may be twice the unemployment rate. However, the countries without these kinds of laws have rates of four or five percent.

Germany issued a law called co-determination which demanded the businesses' decisions, arrangements and work conditions should be decided and agreed both by the employees and the employers. Under this law, the businesses could not fire any workers unless it was agreed to by the latter. Generally speaking, the workers could not be fired after they were 40 years old. Eventually, Germany's labor market lacked of competitive strength, hurting the economy.

The Industrial Disputes Act (IDA) of India, passed in 1947, set the form of mediation between employers and employees. The law also states that when a business wants to fire more than 100 employees, it needs government approval. Businesses in India found it difficult to fire workers after the law came into effect.

To sum up, the study of employment contract is still in progress in the world, and no strong results or explicit implications are made.